\documentclass{article}

\usepackage{graphicx}
\usepackage{psfig}
\usepackage{epsfig}
\usepackage[round]{natbib}

\title{Do probabilistic medium-range temperature forecasts need to allow for non-normality?}

\author{Stephen Jewson\footnote{\emph{Correspondence address}: RMS, 10 Eastcheap,
London, EC3M 1AJ, UK. Email: \texttt{x@stephenjewson.com}}\\
RMS, London, United Kingdom}
\begin{document}

\newcommand{\bx}[1]{\fbox{\begin{minipage}{15.8cm}#1\end{minipage}}}

\maketitle

\begin{abstract}

The gaussian spread regression model for the calibration of site specific ensemble temperature 
forecasts depends on the apparently restrictive assumption that the uncertainty around temperature
forecasts is normally distributed.
We generalise the model using the kernel density to allow for much
more flexible distribution shapes. However, we do not find any meaningful improvement in the resulting
probabilistic forecast when evaluated using likelihood based scores.
We conclude that the distribution of uncertainty is either very close to normal, or if
it is not close to normal, then the non-normality is not being predicted by the
ensemble forecast that we test.

\end{abstract}

\section{Introduction}
\label{introduction}

We consider the question of how to make probabilistic forecasts of
temperature at individual spatial locations. 
Our method is
to take the output from a numerical model ensemble forecast as a predictor,
and to use a statistical model to convert this predictor into a
prediction of the likely temperature distribution. Our interest
is in predictions of the whole distribution of temperatures,
rather than more restricted questions such as whether the
temperature will cross certain thresholds or whether the
temperature will lie in certain categories.

In two previous articles (\citet{jewsonbz03a} and~\citet{jewson03g})
we have addressed this question
using a statistical model that we will call the \emph{gaussian spread
regression} model. This model considers the conditional 
temperature distribution to be normal, and models the observed temperature in
terms of the mean and the standard deviation of the ensemble
forecast. The mean of the observed temperature distribution is taken to be a
linear transformation of the ensemble mean, and the standard
deviation of the observed temperature distribution is taken to be a linear
transformation of the ensemble standard deviation. This model can
be considered as a generalisation of standard linear regression:
the extension is that a linear model, rather than just a constant,
is used for the standard deviation.

There are a number of possible criticisms of the two papers cited above.
In particular:
\begin{itemize}
    \item the autocorrelations of the forecasts errors were ignored when
    fitting the model
    \item the model ignores any predictive information that might be present
    in the previous day's temperatures
    \item the model assumes that temperature uncertainty (the conditional distribution
    of temperature) is normally distributed 
\end{itemize}

We are continuing our research in an attempt to address each of
these issues. This article tackles the third:
the assumption of normality of the conditional temperature distribution.
We present a new model which
no longer makes this assumption. As
with the gaussian spread regression model, this model calibrates the ensemble 
using linear transformations of the ensemble mean
and the ensemble standard deviation. The difference is that the observed
temperatures are then modelled using a \emph{kernel density} rather than a
normal distribution. Kernel densities allow much more flexible
modelling of the probability density of the temperature, and can
incorporate skewness and bimodality, both of which may exist in
the distribution of forecast temperature uncertainty.
We will call the new model \emph{kernel spread regression}.

In~\cite{jewson03g} we showed that there is a detectable relationship
between spread and skill in one year of past forecasts from the ECMWF model for
London Heathrow.
We also showed 
that the gaussian spread regression model applied to these forecasts
performs better than linear regression for in-sample tests. However, in out
of sample tests we could not show that the model performed better. We argue that
this is because one year of data is rather little for fitting the model parameters
accurately, combined with the fact that 
the predictable variations in uncertainty are rather weak.

For these reasons, we see little point in performing out
of sample tests on the models we present below. The currently
available data is simply too short to be able to distinguish
between these models in such tests, since the results of the models
are only subtly different.
We will therefore
only perform in-sample tests and the goodness of the model
will be assessed on the basis of in-sample results. 
Testing using out of sample tests will have to
wait until longer time series of stationary past forecasts are
available.

In section~\ref{data} we will describe the forecast data to be used
in this study. In section~\ref{kernel} we introduce the kernel
density, and present some properties of kernel densities that will be useful in
interpreting the results of our analysis. 
In section~\ref{model} we present the hierarchy of statistical calibration models
that we will consider. The models are: gaussian regression, gaussian spread regression,
kernel regression and kernel spread regression. 
In section~\ref{results} we present results from the use of these four models to 
calibrate some forecast data. Finally in section~\ref{conclusions} we summarise our results,
discuss areas for future work, and draw some conclusions.

\section{Data}
\label{data}

We will base our analyses on one year of ensemble forecast data for the weather
station at London's Heathrow airport, WMO number 03772. The forecasts are predictions
of the daily average temperature, and the target days of the forecasts
run from 1st January 2002 to 31st December 2002. The forecast was produced
from the ECMWF model~\citep{molteniet96} and downscaled to the airport location using a simple
interpolation routine prior to our analysis. There are 51 members in the ensemble.
We will compare these forecasts to the quality controlled climate
values of daily average temperature for the same location as reported by the UKMO.

There is no guarantee that the forecast system was held constant throughout this period,
and as a result there is no guarantee that the forecasts are in any sense stationary,
quite apart from issues of seasonality. This is clearly far from ideal with respect to 
our attempts to build statistical interpretation models on past forecast data but is,
however, unavoidable: this is the data we have to work with.

Throughout this paper all equations and all values have had both the seasonal mean and
the seasonal standard deviation removed. Removing the seasonal standard deviation
removes most of the seasonality in the forecast error statistics, and partly justifies the use of
non-seasonal parameters in the statistical models for temperature that we propose.

\section{The kernel density}
\label{kernel}

Two of the four models that we consider in this study make use of 
kernel densities. 
The kernel
density is a very flexible method for deriving a density function
from observed data. It works by assigning an identical 
kernel to each individual observed data point. The fitted density is then
the sum of these individual kernels. There is a single free
parameter, which is the width of the kernel, known as the
\emph{bandwidth}. The kernels are normalised so that the total
density is one. A number of different kernel shapes are possible,
but for simplicity, and to ensure that our densities never give
zero probabilities, we will use gaussian kernels. This gives a
probability density of the form:

\begin{eqnarray}
  p(x)&=&\sum_{i=1}^{i=M} p_i(x)\\\nonumber
      &=&\sum_{i=1}^{i=M} \frac{1}{\sqrt{2\pi} \lambda M} \mbox{exp} \left( -\frac{(x-x_i)^2}{2 \lambda^2 } \right)  
\end{eqnarray}

where $\lambda$ is the bandwidth, $M$ is the number of data
points, and $x_i$ are the data points themselves.
In our case the data points $x_i$ will be the individual ensemble members,
after certain transformations that are described in section~\ref{model}.

We now derive an
important relationship which links the sample variance of the
data points $x_i$ to the variance of the fitted kernel distribution $p(x)$.

If $x$ is the temperature, then the expectation of $x$ calculated using the probabilities from
the calibrated forecast $p(x)$ is:
\begin{equation}
  \mu=\int x p(x) dx
\end{equation}
But $p(x)$ is given by a kernel density, and so:
\begin{equation}
  p(x)=\sum_{i=1}^{i=M} p_i(x) 
\end{equation}
where
\begin{equation}\label{norm}
  \int p_i(x) dx=\frac{1}{M}
\end{equation}

and so:
\begin{eqnarray}
  \mu&=&\int x p(x) dx \\\nonumber
     &=&\int x \left( \sum_{i=1}^{i=M} p_i(x) \right) dx \\\nonumber
     &=&\sum_{i=1}^{i=M} \int x  p_i(x) dx \\\nonumber
     &=&\frac{1}{M} \sum_{i=1}^{i=M} \int x  M p_i(x) dx     
\end{eqnarray}

The normalisation of $p_i(x)$ (equation~\ref{norm}) means that we can consider
$M p_i(x)$ to be a density with mean of $x_i$, where
$x_i$ is the $i$'th ensemble member, and so:

\begin{equation}
  \mu=\frac{1}{M} \sum_{i=1}^{i=M} x_i
\end{equation}

In other words, the mean of $x$ under the density $p(x)$ is just the arithmetic
mean of the ensemble members $x_i$. 

Similarly we can consider the variance of $x$ under the density $p(x)$:

\begin{eqnarray}
 var(x)&=&\int (x-\mu)^2 p(x) dx \\\nonumber
       &=&\int (x-\mu)^2 \left( \sum_{i=1}^{i=M} p_i(x) \right) dx \\\nonumber
       &=&\sum_{i=1}^{i=M} \int (x-\mu)^2 p_i(x) dx \\\nonumber
       &=&\sum_{i=1}^{i=M} \int (x-x_i+x_i-\mu)^2 p_i(x) dx \\\nonumber
       &=&\sum_{i=1}^{i=M} \int [(x-x_i)^2+(x_i-\mu)^2+2(x-x_i)(x_i-\mu)] p_i(x) dx \\\nonumber
       &=& \frac{1}{M} \sum_{i=1}^{i=M} \int (x-x_i)^2 M p_i dx
          +\frac{1}{M} \sum_{i=1}^{i=M} \int (x_i-\mu)^2 M p_i dx
          +\frac{2}{M} \sum_{i=1}^{i=M} (x_i-\mu) \int (x-x_i) p_i dx 
\end{eqnarray}

The summand in the first of these terms is just the variance of $x$ under the density $Mp_i$. 
This is equal to $\lambda^2$, since
$Mp_i$ is a normal distribution with variance $\lambda^2$.
In the second term, we note that $\int M p_i(x) dx=1$.
In the third term, we note that $\int (x-x_i)M p_i(x) dx=0$ by the definition of $x_i$.

Putting this all together:

\begin{equation}
 var(x)=\lambda^2
         +\frac{1}{M} \sum_{i=1}^{i=M} (x_i-\mu)^2 \\\nonumber
\end{equation}

But the second of these terms is just the sample variance of the ensemble members, and so we have:
\begin{equation}
\label{variance}
  \mbox{variance of modelled temperatures}=\lambda^2+\mbox{sample variance of ensemble members}
\end{equation}

What we see is that the variance of the fitted density is equal to
the sample variance plus the bandwidth squared. This relation will
help us reconcile results from the gaussian and kernel based
models given below.

\section{The kernel spread regression calibration model}
\label{model}

We now describe each of the models we will use in this study in
turn. These models are: gaussian regression, 
gaussian spread regression, kernel regression and kernel spread regression. We will use these
four models to compare the benefits of using the ensemble spread as
a predictor with the benefit of allowing non-normal, rather than just normal,
forecast distributions.

\subsection{Gaussian regression}

The simplest model we will use for observed temperatures is just
linear regression: we will call this \emph{gaussian regression} to
distinguish it from the kernel regression model described below, 
which replaces the normal distribution with a kernel density. 
We will write the gaussian regression model as:

\begin{equation}
 T_i=N(\alpha+\beta m_i, \gamma)
\end{equation}

$\alpha, \beta, \gamma$ are the parameters of the model, and $m_i$
is the ensemble mean. This model, and its performance on this data
set, has been discussed extensively in~\citet{jewson03g}.

\subsection{Gaussian spread regression}

The second model we will use to model the observed temperatures is
the gaussian spread regression model of~\citet{jewsonbz03a}. We will write this
model as:

\begin{equation}
 T_i=N(\alpha+\beta m_i, \gamma+\delta s_i)
\end{equation}

Relative to the gaussian regression model there is an extra
parameter $\delta$ which scales the extra predictor $s_i$ 
(the ensemble standard deviation). The performance of this model,
and the interpretation of the parameters, has been discussed
extensively in~\citet{jewsonbz03a} and~\citet{jewson03g}. 

The gaussian regression and gaussian spread regression models are included 
here as a basis for comparison for the kernel based models.

\subsection{Kernel regression}

The first of the two new models that we will present here is an adaption
of standard gaussian regression that uses a kernel density rather
than a normal distribution. The motivation for this is that the uncertainty in temperature forecasts
is not necessarily normally distributed (indeed, is
undoubtedly not \emph{exactly} normally distributed) 
and so it does not necessarily make sense to apply a
gaussian model. In the kernel regression model we predict the mean of future
temperatures in exactly the same way as we do with the gaussian
model: our prediction is a linear transformation of the
ensemble mean. The difference from gaussian regression is that the
distribution around the mean is derived from the individual ensemble 
members using the kernel density. To be consistent with this the
parameters of the model are fitted using the likelihood calculated
from the kernel density rather than the normal distribution.

We will write this model as:

\begin{equation}
 T_i=K(\alpha+\beta m_i, \gamma, \lambda)
\end{equation}

where the notation has been chosen to emphasize the similarity
with gaussian regression, but with the extra bandwidth parameter $\lambda$. 
The way this model works in detail is:

\begin{itemize}
    \item The ensemble members are transformed using the linear transformation
    $\alpha+\beta x_i$.
    \item The standard deviation of the ensemble members is set to
    $\gamma$
    \item A kernel of bandwidth $\lambda$ is placed around each of the transformed
    ensemble members
    \item The final density is given by the sum of these kernels
\end{itemize}

We note that this model, along with the gaussian regression model, ignores
any information in the ensemble standard deviation because the standard 
deviation of the members is forced to a fixed value of gamma.

\subsection{Kernel spread regression}

The second of the two new models that we present generalises the
kernel regression model in the same way that gaussian spread
regression generalises gaussian regression. In other words, this model
extends the kernel regression model to include the ensemble spread
as a predictor for the uncertainty of the predicted temperatures.
This model is intended to have all the features of the gaussian
spread regression model, but also to have the advantage that it
can be used for forecast uncertainty that is not close to normally
distributed, perhaps because of skewness or bimodality. We write
this model as:

\begin{equation}
 T_i=K(\alpha+\beta m_i, \gamma+\delta s_i, \lambda)
\end{equation}

\subsection{Parameter fitting}

We fit the parameters of all four of our models by finding those
values that maximise the likelihood of the observations
given the model for our one year of data. 
We perform this fitting using an iterative
optimisation scheme known as Newton's method. For the gaussian
models, this is very fast. For the kernel models it is
significantly slower because evaluation of the probability density
of the models involves the summing over kernels for each of the
ensemble members. However, fitting all 10 leads still only takes a
matter of seconds on a rather slow personal computer (100Mhz CPU).

\section{Results}
\label{results}

We now present the results of fitting our four
models to the forecasts and the observed temperature data. We
start with the results for the gaussian and kernel regression models
and then show results for the spread regression models. 

\subsection{Regression model parameters}

\begin{figure}[!htb]
  \begin{center}
    \includegraphics{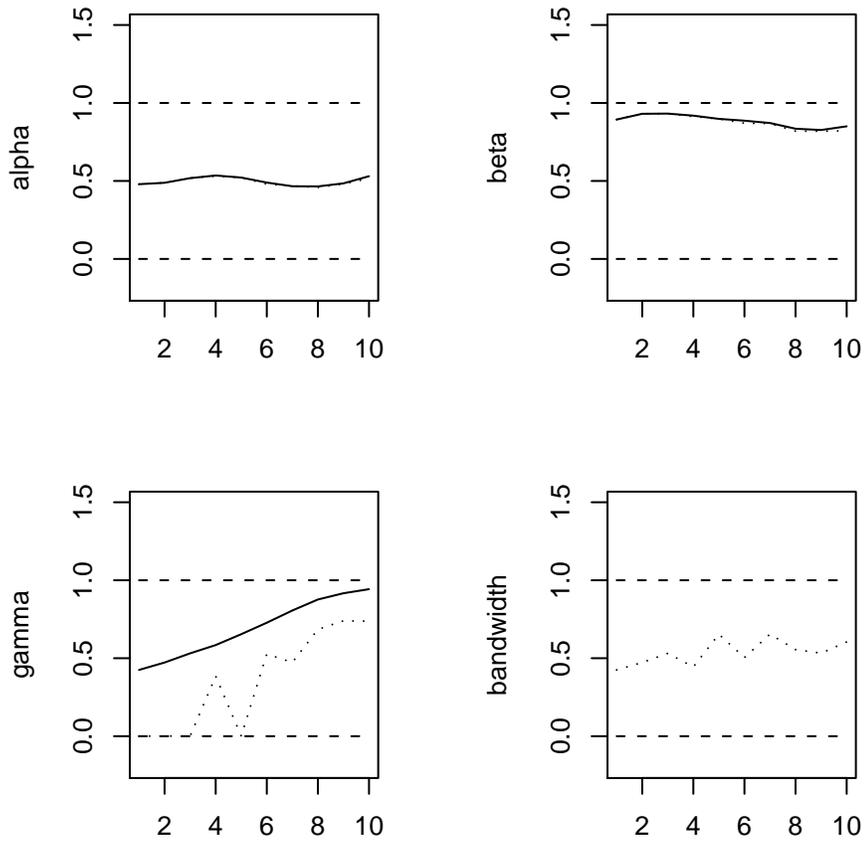}
  \end{center}
  \caption{
 The solid line shows parameter values for the gaussian regression model, 
 and the dotted line shows parameter values for the kernel regression model. 
 The kernel regression has one extra parameter (the bandwidth).
          }
  \label{f:f1}
\end{figure}

Parameters from the fitting of the gaussian and kernel regression models are shown
in figure~\ref{f:f1}. These models share the parameters $\alpha, \beta,
\gamma$, while the kernel regression model has the extra parameter $\lambda$. 
We see that the values of $\alpha, \beta$ are effectively
the same for the two models. 
The $\gamma$ parameter, however, is significantly different: 
it is much lower for the kernel regression than it
is for the gaussian regression. The reason for this seems to lie
in equation~\ref{variance}. In the gaussian model, the variance of
temperature is captured by $\gamma$, while in the
kernel model the variance is captured through a combination of 
$\gamma$ and the bandwidth. Thus the $\gamma$ can be partially
offset by the bandwidth in this model. 
We will explore later, for the spread
regression model, whether the resulting calibrated variances are
the same or not.

That the bandwidth is offsetting the $\gamma$ raises the
possibility that the model is overspecified because these two
parameters are actually trying to capture the same thing, and that
only one would do. It would certainly be possible to fit a kernel
based model that had a \emph{fixed} bandwidth...presumably the
$\gamma$ parameter would then adjust to fit the observed variance.
Similarly it would be possible to fit a kernel model that had a
fixed value for $\gamma$: presumably the bandwidth would then
adjust to fit the observed variance. However, it does seem that
the presence of two independent parameters is beneficial and that
the two parameters can potentially play different roles in the calibration. 
This is because the bandwidth modulates the amount of non-normality.
For instance, the bandwidth can adjust to make the total density more
or less unimodal (large values) or multimodel (small values).

\subsection{Spread regression model parameters}

\begin{figure}[!htb]
  \begin{center}
    \includegraphics{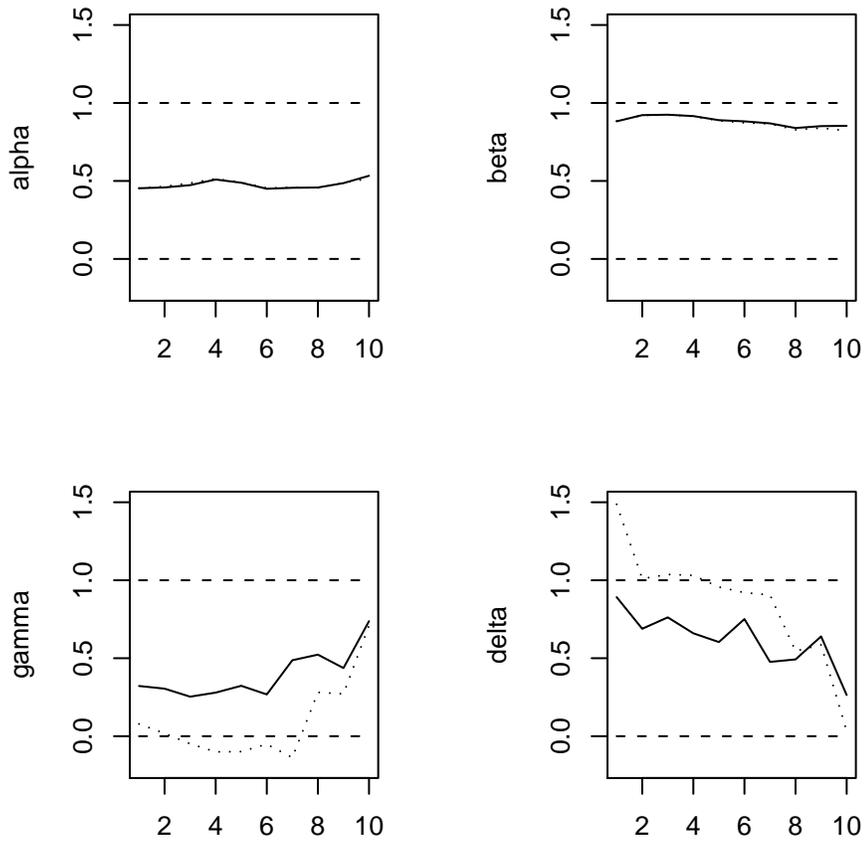}
  \end{center}
  \caption{
  The values for the parameters $\alpha,\beta,\gamma,\delta$ for the gaussian spread
  regression model (solid line) and kernel spread regression model (dotted line).
          }
  \label{f:f2}
\end{figure}

We now consider the parameters from the gaussian and kernel spread regression
models. Figure~\ref{f:f2} shows that the values of $\alpha$ and $\beta$ are
effectively the same as for the regression models. As with the
regression models, the value of $\gamma$ for the kernel model is
much lower than that for the gaussian model. The value of $\delta$
is also different between the two models. The extent to which the
different values of $\gamma$ and $\delta$ lead to the same or different final
levels for the mean and variability of the calibrated spread is
discussed below.

\begin{figure}[!htb]
  \begin{center}
    \includegraphics{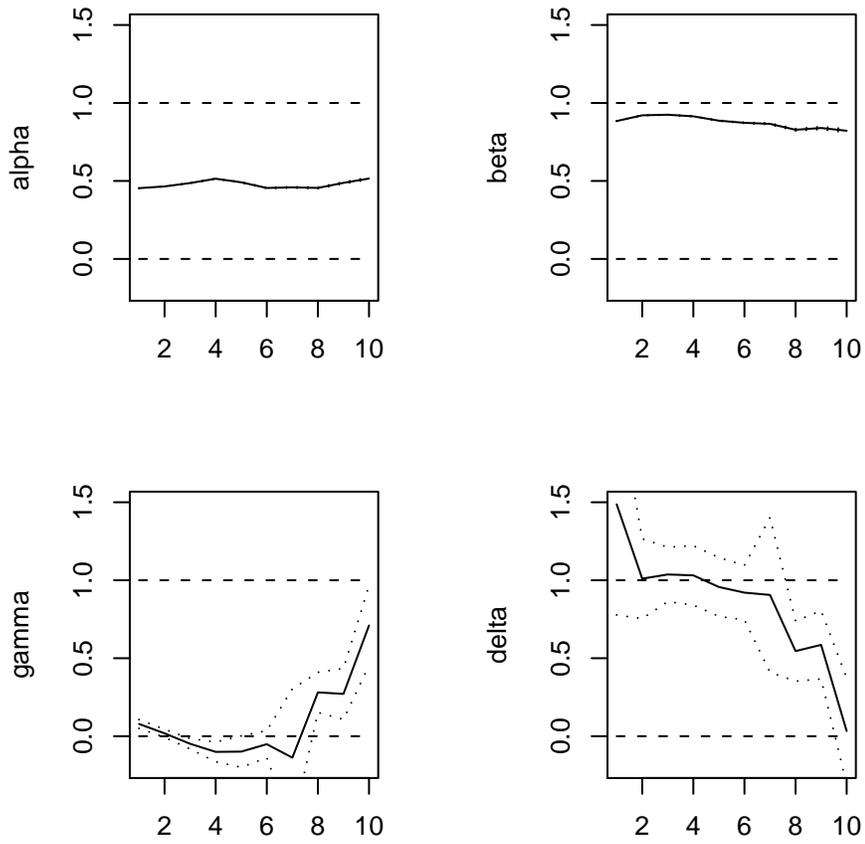}
  \end{center}
  \caption{
     The $\alpha, \beta, \gamma, \delta$ parameters for the kernel spread
     regression model, with confidence intervals.
          }
  \label{f:f2b}
\end{figure}

Figure~\ref{f:f2b} shows the parameters for the kernel spread regression model alone,
but with estimated confidence intervals (estimated in the standard way using the 
Fisher information).
We see that the parameters for the ensemble mean
are very well estimated 
(so much so that the confidence intervals are almost hidden by the data itself) 
while the parameters for the standard deviation are much
more uncertain, particularly the delta parameter. 

\begin{figure}[!htb]
  \begin{center}
    \includegraphics{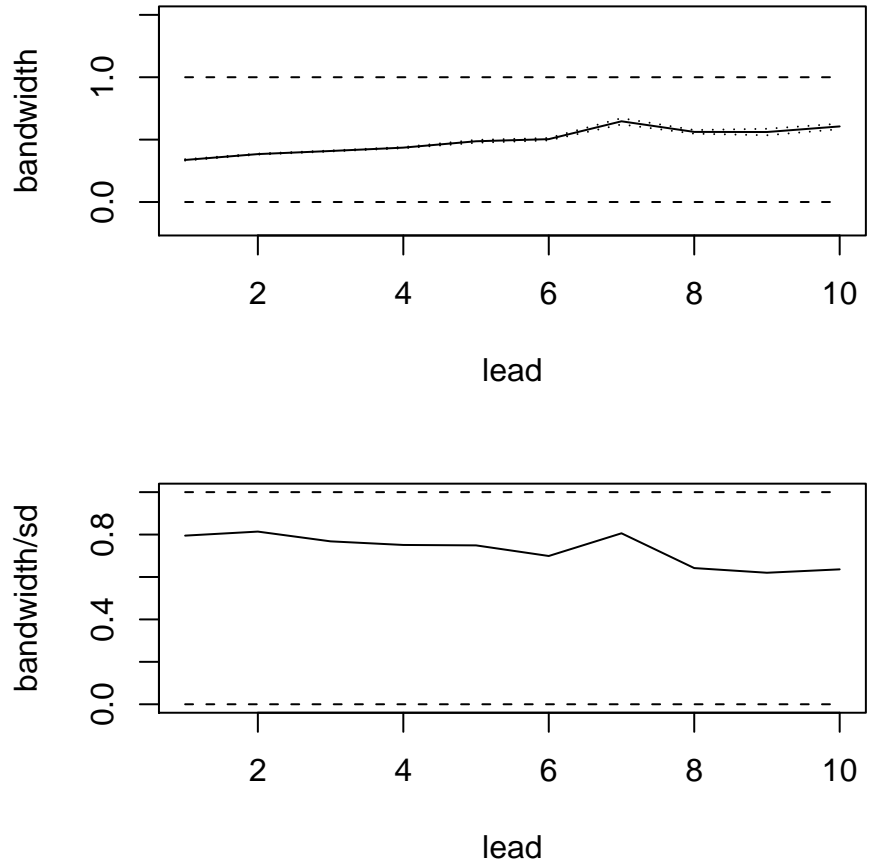}
  \end{center}
  \caption{
  The upper panel shows the bandwidth for the kernel spread regression model, with confidence
  intervals, and the 
  lower panel shows the ratio of this bandwidth to the standard deviation after calibration.
          }
  \label{f:f3}
\end{figure}

The value of $\lambda$ for the spread regression model is shown in the upper panel of
figure~\ref{f:f3}, along with confidence intervals. This parameter is apparently
very well estimated. 

\subsection{Density examples}

\begin{figure}[!htb]
  \begin{center}
    \includegraphics{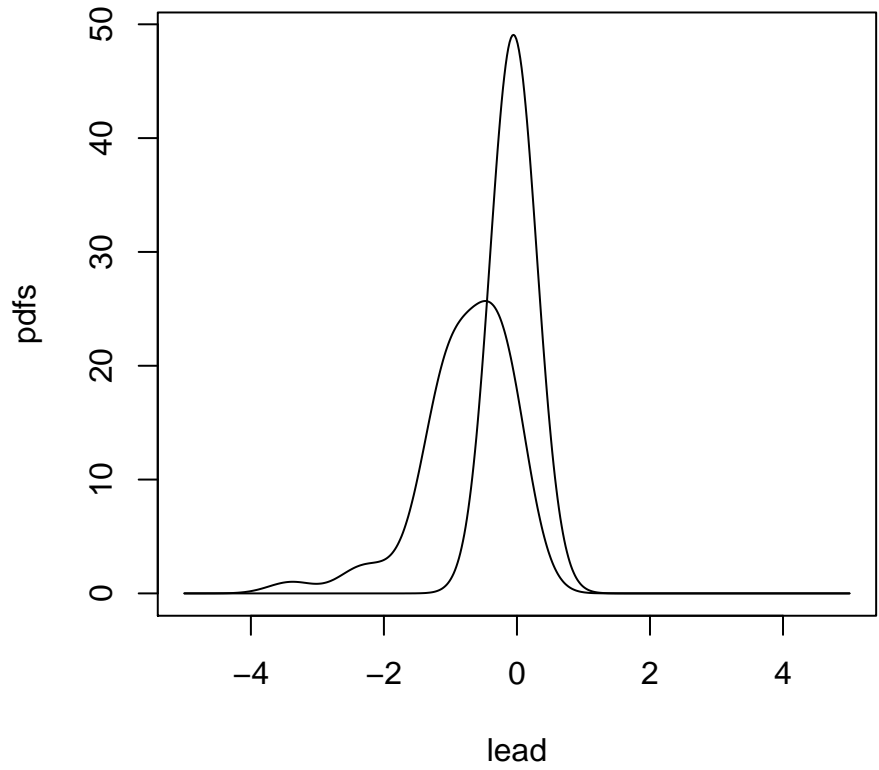}
  \end{center}
  \caption{
   The two curves show the densities predicted by the kernel spread regression model at lead 10 
   on the days of the year with the highest and the lowest ensemble standard deviations. 
          }
  \label{f:f7}
\end{figure}

For illustration we now show, in figure~\ref{f:f7}, 
two examples of the densities that are produced by the kernel spread
regression. These examples are both from lead 10, and
show the densities predicted by the model on the days
with the highest and lowest ensemble standard deviations.
The same bandwidth $\lambda$ is applied to these two cases by the model, and
so we would expect that the high standard deviation case might show some
multimodality, while the low standard deviation case is unlikely to. This is 
exactly what we see. We note, however, that the presence of multimodality
in the wider of the two densities is not at all an indication that we have detected multimodality in the
forecast: it is simply the output of the model on this particular day. The optimum value
of the bandwidth is based on all days, and the optimisation process used for fitting the
parameters may have 'sacrificed' the goodness of fit on this particular day in order to 
benefit the likelihood score based on the entire data set. 

\subsection{Non-normality indicator}

If the distribution of forecast uncertainty is frequently significantly non-normal one would 
expect the bandwidth in the kernel spread regression model 
to be small relative to the total standard deviation of the calibrated forecast.
The lower panel of figure~\ref{f:f3} shows the ratio of these two terms. In fact, this ratio
is never small, indicating only a low level of non-normality is present in our data-set. The decrease
in this ratio with lead time suggests that the non-normality increases slightly at longer leads,
as one would expect. 

\subsection{Statistics of the calibrated spread}

\begin{figure}[!htb]
  \begin{center}
    \includegraphics{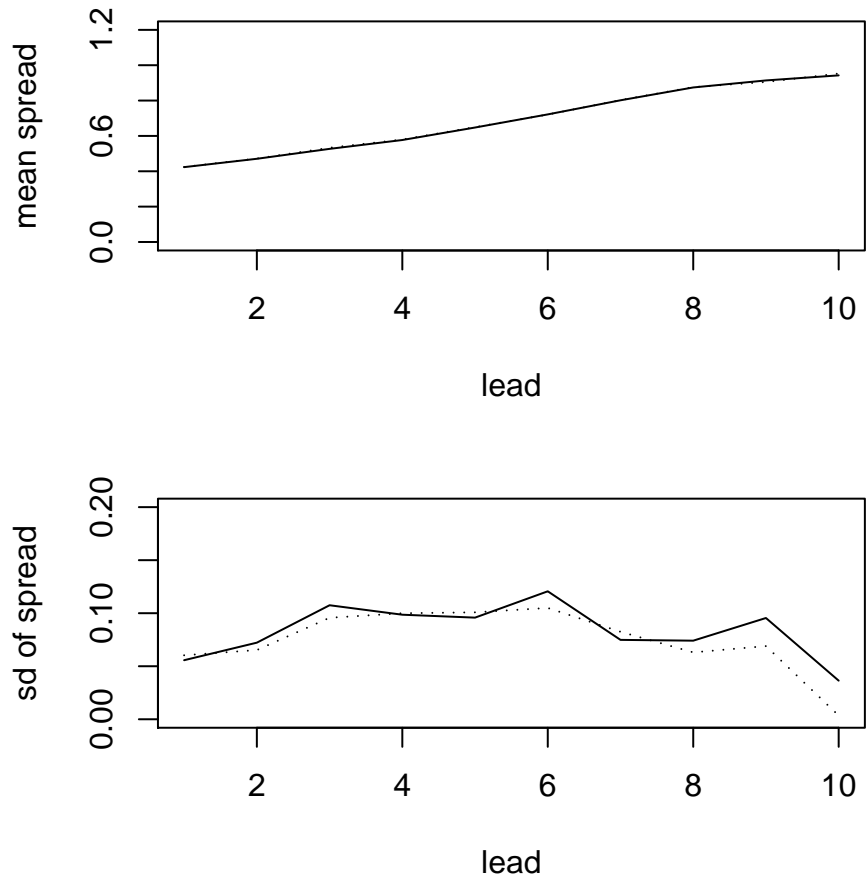}
  \end{center}
  \caption{
  The upper panel shows the mean uncertainty predicted by the gaussian spread regression model (solid line)
  and the kernel spread regression model (dotted line). The lower panel shows the standard deviation of 
  the uncertainty for the same two models.
          }
  \label{f:f5}
\end{figure}

We now look at the statistics of the uncertainty of the calibrated forecast predicted by the
two spread regression models. In particular, we consider the mean
level of the uncertainty, and the standard deviation of the
uncertainty. These are both shown in figure~\ref{f:f5}. 
In the upper panel we see that the
mean level of the uncertainty in the two models is effectively the
same (the lines are virtually indistinguishable). 
Even though the values of $\gamma$ are different, the value
of $\lambda$ in the kernel spread regression model is exactly
compensating so that the two models produce the same mean level of
uncertainty. We also see that the same is broadly true for the
variability of the uncertainty (lower panel) although with some small
differences between the models. We suspect that these differences
are simply due to sampling errors: fitting the $\delta$ parameter
is much more difficult than fitting the $\gamma$ parameter because
it is associated with the second moment of the second moment of
the temperatures. We conclude that even though the values of
$\delta$ are different in the two models, the combined effect on
the variability of the calibrated spread is the same.

\subsection{Likelihood scores}

\begin{figure}[!htb]
  \begin{center}
    \includegraphics{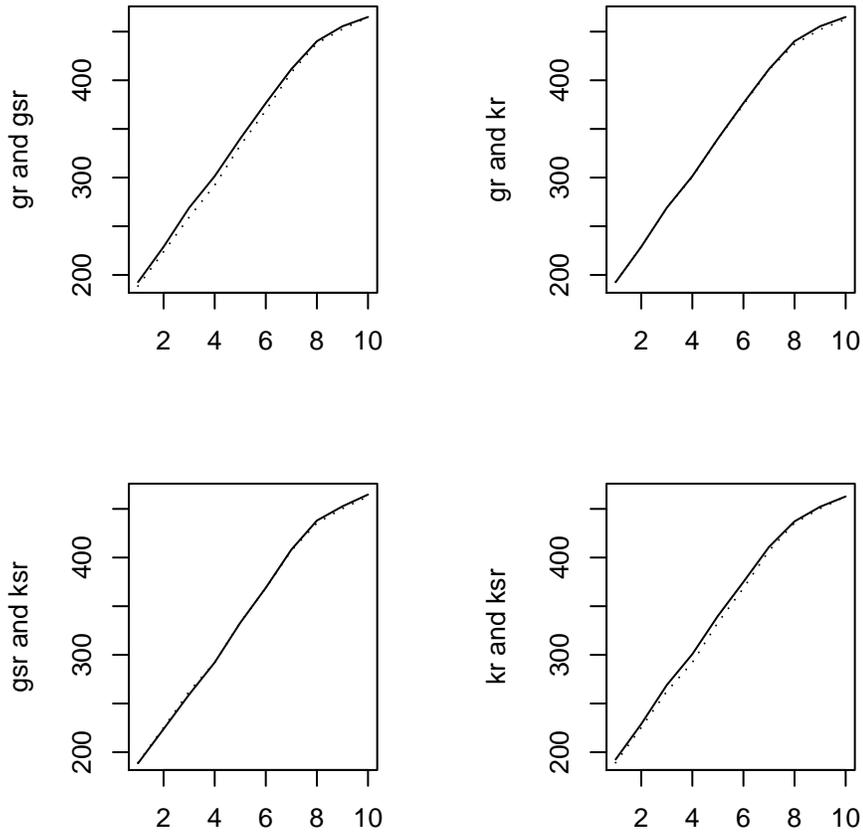}
  \end{center}
  \caption{
     The four panels show the results of comparing the MLL (minus log likelihood) skill measures for the
     four models (low values are better).
     The upper left panel compares gaussian regression with gaussian spread regression. 
     The upper right panel compares gaussian regression with kernel regression.
     The lower left panel compares gaussian spread regression with kernel spread regression and
     the lower right panel compares kernel regression with kernel spread regression. 
          }
  \label{f:f6}
\end{figure}

Our final set of results, in figure~\ref{f:f6}, compare the performance of
the various models using the log-likelihood. This tells us which of
the models does best in representing the observed temperatures, and
is the acid-test of whether modelling non-normality confers any benefit.
Because the various models do not have the same number of
parameters, we have corrected the log-likelihoods using the BIC
criterion (although the differences in the numbers of parameters is so small
relative to the amount of data this makes no qualitative difference).

The likelihoods from the spread regression models are slightly
better than those from the regression models. This is presumably
because the uncertainty in the temperature forecast shows a small
amount of flow dependence, and this flow dependence has been
partly predicted by the forecast model. This issue is discussed in
more detail in~\citet{jewson03g}.

If the distribution of prediction uncertainty were significantly non-normal
(e.g. skewed or multimodal) and the forecasts were predicting this non-normality, 
we
might expect that the likelihood values for the kernel models
would be significantly better than those for the gaussian models.
However, we do not see this in the results. In fact, the differences
between the gaussian and kernel models are tiny.
This suggests that
non-normality in the distribution of forecast uncertainty 
is either not particularly strong, or that the forecasts are not
capturing it (although we cannot distinguish between these two 
possibilities).

On very close inspection, we see that the log-likelihoods for the
kernel models are \emph{slightly} better than those for the
gaussian models at long leads. 
This corresponds to the idea that non-normality in the distribution of
forecast uncertainty is
more likely at long leads, as non-linear processes will have had
more time to take effect. However, a \emph{much} longer data set of past
forecasts would be needed to show that this effect is not just an
artefact of sampling error. In any case, the difference would seem to
be too small to be important.

\section{Conclusions}
\label{conclusions}

We have extended both linear regression and the gaussian spread regression model of~\citet{jewsonbz03a} 
to cope with more general distribution shapes by using kernel
densities instead of the normal distribution. 
This gives us a hierarchy of four models: gaussian regression,
gaussian spread regression, kernel regression and kernel spread regression. We apply all four
models to the calibration of one year of ensemble forecasts. 
The parameters
of the linear transformation of the ensemble mean 
are the same in all the models.
However, the parameters of the
transformation of the standard deviation of the ensemble are different.
This appears to be because part of the variance of the observed
temperatures is taken up by the bandwidth parameter of the
kernel models. 
Nevertheless, although the parameter values change, the mean and standard
deviation of the final calibrated uncertainty is more or less the same
in the gaussian and kernel models.

We compare the likelihoods attained by the four models. 
The spread regression models perform marginally better than the regression models.
But the kernel models do not perform better than the gaussian models, except
for \emph{very slightly} higher likelihoods at long leads. 
We conclude that either the forecast uncertainty
is not non-gaussian, or if it is, then the forecast does not capture that. 
Any non-normality in the forecast ensemble does not improve the forecasts, and 
can be ignored. 

Which of the gaussian and kernel models should be used in
practice? The gaussian models are certainly the simpler of the two and
only involve manipulating the ensemble mean and standard
deviation rather than ensemble members. They are easier and faster to
fit, and the final distribution is easier to communicate as it can
be summarised by the mean and the standard deviation. However, the
kernel models are more general. We have only tested one station:
there may be other locations for which the kernel model shows a greater
benefit, and they are unlikely to do worse.
These arguments are not conclusive in either direction, and
which model to use seems to be a 
matter of personal preference at this point \footnote{We prefer the gaussian models, 
because of their simplicity.}.
Only the availability of
longer stationary time series of past forecasts could 
allow us to come to a firm conclusion as to whether one model is really better than the other.

There are still some aspects of the calibration of temperature
forecasts that need investigation. One is the need for better
modelling of the covariance matrix of forecast errors. \citet{jewson03g} showed
that the residuals from the gaussian spread regression model are
certainly not white in time, and this should really be taken into account
in the model: this could possibly affect the results
significantly. We note that the modelling of such autocorrelated residuals would
be much easier in the context of the gaussian 
models than the kernel models.

The models we have considered are parametric.
It is possible that non-parametric calibration models may also be
possible, and may have some advantages over their parametric
cousins. This would be worth investigating. However, the main benefit
of non-parametric models is in the flexible range of distributions
that they can represent. Given that our results show no material benefit from
reducing the dependence on the normal distribution and using a much more 
flexible distributional shape, we doubt that non-parametric models would help either.

Finally, we mention the question of extending the kinds of
calibration methods and models used in this study to other
variables such as wind and precipitation. Neither wind nor
precipitation are close to normally distributed, and it would not be
appropriate to try and calibrate them using the gaussian 
models. However, for wind at least, it may be possible
that the kernel models would work quite well. We
plan to test this just as soon as we can get our hands on some
ensemble wind forecasts and wind observations.

\newpage

\newpage

\newpage

\newpage


\newpage

\newpage

\newpage

\bibliography{kernelspreadregression}

\end{document}